\documentclass[conference]{IEEEtran}
\IEEEoverridecommandlockouts
\usepackage{cite}
\usepackage{amsmath,amssymb,amsfonts}
\usepackage{algorithm}
\usepackage{algorithmic}
\usepackage[utf8]{inputenc}
\usepackage{multirow}
\usepackage{array}
\usepackage{graphicx}
\usepackage{lipsum}
\usepackage{subcaption}
\usepackage{textcomp}
\usepackage{xcolor}
\usepackage{amsmath}
\usepackage[left=0.625in, right=0.625in, top=0.75in, bottom=1in, footskip=0.25in]{geometry}
\usepackage{tikz}
\usepackage{amssymb}
\usepackage{graphicx}
\usepackage{subcaption}
\usepackage{cite}
\usepackage{bm}
\usepackage{dsfont}
\usepackage{graphics} 
\usepackage{epsfig} 
\usepackage{stackrel}
\usepackage{geometry}{}
\usepackage{float}
\usepackage{url}
\def\BibTeX{{\rm B\kern-.05em{\sc i\kern-.025em b}\kern-.08em
    T\kern-.1667em\lower.7ex\hbox{E}\kern-.125emX}}
\usepackage{ifthen}

\begin{document}

\title{\textbf{RAIL}: \textbf {R}obust \textbf {A}coustic \textbf {I}ndoor \textbf {L}ocalization for Drones}
\author{\IEEEauthorblockN{Alireza Famili, Angelos Stavrou, Haining Wang, Jung-Min (Jerry) Park}
\IEEEauthorblockA{\textit{Department of Electrical and Computer Engineering, Virginia Tech} \\
\{afamili, angelos, hnw, jungmin\}@vt.edu}
}

\maketitle

\begin{abstract}
Navigating in environments where the GPS signal is unavailable, weak, purposefully blocked, or spoofed has become crucial for a wide range of applications. A prime example is autonomous navigation for drones in indoor environments: to fly fully or partially autonomously, drones demand accurate and frequent updates of their locations. This paper proposes a Robust Acoustic Indoor Localization (RAIL) scheme for drones designed explicitly for GPS-denied environments. Instead of depending on GPS, RAIL leverages ultrasonic acoustic signals to achieve precise localization using a novel hybrid Frequency Hopping Code Division Multiple Access (FH-CDMA) technique. Contrary to previous approaches, RAIL is able to both overcome the multi-path fading effect and provide precise signal separation in the receiver. Comprehensive simulations and experiments using a prototype implementation demonstrate that RAIL provides high-accuracy three-dimensional localization with an average error of less than $1.5$~cm.
\end{abstract}


\section{Introduction} \label{Introduction}
\noindent Over the past few years, the global drone industry has expanded exponentially and the number of use cases in which drones play a significant part, both in indoor and outdoor environments, has flourished. Indeed, there is a wide range of indoor drone deployment nowadays, ranging from recreational use to life-saving matters. There are a plethora of representative examples for drones that range from reconnaissance inside hazardous and shielded facilities (e.g., nuclear power plants), assisting first responders in navigating inside buildings and confined spaces, to inventory maintenance and security surveillance inside large warehouses among others~\cite{Matthan,ROLATIN}.

In most of the aforementioned examples, drones must have full or partial autonomous flying capabilities to perform their tasks successfully. To achieve any degree of flight autonomy between the current position and the target destination, the drone's navigation system needs to have access to accurate and frequent localization information. In outdoor environments, there is usually access to GPS signals for self-localization; however, GPS-assisted navigation is not reliable or even available in indoor spaces. Moreover, there are many GPS-denied areas where the GPS signal is blocked or purposefully spoofed.

In absence of reliable GPS signals, vision-based methods are widely used for localization and navigation of drones~\cite{Novel_Visual_Odometry}. However, the accuracy of current vision-based approaches is usually limited due to the drone's vibration during flight. In addition, the location accuracy can degrade even further in vision-impaired environments. Moreover, vision-based methods have high computational complexity and deployment cost rendering them impractical for small indoor drones that require frequent location updates and low energy sensors to operate. In addition to vision-based approaches, ranging-based methods are commonly deployed for indoor localization. Chief among the latter approaches are ones that employ Radio Frequency (RF)~\cite{Freq_Hopping_WiFi} or acoustic~\cite{ROLATIN} signals. Unfortunately, the performance of these methods is significantly degraded in indoor environments due to multi-path fading~\cite{ROLATIN}.

To address accuracy degradation due to multi-path fading, we propose \emph{RAIL (Robust Acoustic Indoor Localization)}, a three-dimensional positioning scheme for autonomous drones in GPS-denied environments. RAIL uses ultrasonic acoustic-based signals for localization. Acoustic signal localization approaches have some advantages over RF-based ones. Most importantly, the significantly slower propagation speed of the acoustic signals allows for higher accuracy with considerably cheaper equipment. In addition, RF signals can penetrate through room boundaries causing interference errors to the localization measurement. Moreover, there are some places where the RF signals' deployment is banned due to security issues. All being said, RAIL uses high-frequency acoustic signals, known as \emph{ultrasounds}, to prevent any interference with human-generated or drone's propeller noise. 

RAIL employs a hybrid \textit{Frequency Hopping Code Division Multiple Access (FH-CDMA)} waveforms to provide a multi-path-robust ranging and significantly mitigate the localization error. In addition, it provides signal separation at the receiver with low latency. \textit{Frequency Hopping (FH)} is a well-known technique that has been mainly used in military communications as an anti-jamming strategy or in Bluetooth technology to avoid interference with co-existing WiFi channels. On the other hand, \textit{Code Division Multiple Access (CDMA)} is a signal spreading technique that has been primarily used in third-generation cellular communication to provide multi-user capability. To the best of our knowledge, RAIL is the first to propose employing the hybrid FH-CDMA for a completely different and novel use-case: providing a multi-path-robust ultrasonic ranging for a seamless three-dimensional localization of drones in GPS-denied areas. Following is a summary of our contributions.

$\bullet$ We propose RAIL, a novel three-dimensional localization scheme for drones in GPS-denied environments which is robust against multi-path fading and provides location estimation with high accuracy.

$\bullet$ RAIL is the first scheme to employ the hybrid \textit{FH-CDMA} for ranging to make it resilient against multi-path fading and provide signal separation at the receiver.

$\bullet$ RAIL reduces the communication exchanges required for navigation by placing the receiver on-board the drone and transmitter beacons in the room. This becomes feasible due to the use of the CDMA techniques.

$\bullet$ Our simulation and experimental results indicate that RAIL's localization error is low, approximately $1.5$ centimeters on average, in three-dimensional space across different trajectories.

The rest of this paper is organized as follows. In the next section, we briefly review the related work. Then, in Section~\ref{Robust FH-CDMA Localization}, we explain how RAIL works in GPS-denied environments. In Section~\ref{Simulation}, we detail our simulation setup and results. Our experimental testbed and outcomes from our prototype are discussed in Section~\ref{Experimental Tests and Results}. Finally, we conclude our work in Section~\ref{Conclusion}.

\section{Related Work} \label{Related Work}
Pinpointing a moving target in indoor environments without the GPS signal has been a topic of interest. Ranging-based methods are of the most well-known approaches for indoor localization. In this category RF, acoustic, or ultrasound signals are employed to find the distance between the beacons and the target. By combining distance measurements between the target and several beacons, the target's position is estimated using lateration or angulation~\cite{ROLATIN,CAT,MobiSys_Follow_Me_Drone,ToneTrack,Robust_Broadband}. However, in indoor environments, traditional ranging techniques face performance challenges due to noise and multi-path fading~\cite{ROLATIN,MobiSys_Follow_Me_Drone}.

Another class of approaches involves vision-based models. Using different visual information processing such as visual odometry (VO), simultaneous localization and mapping (SLAM), and optical flow \cite{low_cost_solution,6_Dimensional,Survey_UAV_navigation_GPS_denied}. There are also some research papers where they used deep neural networks in combination with visual techniques~\cite{Neural_Network} or use of LiDAR~\cite{LiDAR} for autonomous flying. However, all of these vision-based techniques require costly sensors and extensive computing resources and energy that make them impractical for indoor drone use.

\section{Robust Localization with Hybrid FH-CDMA Ultrasound Signals} \label{Robust FH-CDMA Localization}
RAIL is a novel and highly accurate three-dimensional localization scheme for drones designed for indoor environments. RAIL uses the hybrid FH-CDMA technique to overcome the multi-path effects and improve the accuracy of localization. It also harnesses an additional ultrasonic range measurement sensor to compensate for the $Z-$axis estimation error due to the relative geometry between the transmitters and the receiver. 

This section thoroughly investigates how RAIL provides a high-accuracy three-dimensional localization by making the system robust against noise and the indoor multi-path fading effects.

\subsection{Measurement Methods and Location Estimation Techniques}
RAIL leverages ultrasonic acoustic signals for distance estimation because they have advantages over the other ranging techniques, as we discussed in Section~\ref{Introduction}. Well-known measurement methods for ranging are the angle of arrival (AOA), time of arrival (TOA), time difference of arrival (TDOA), and received signal strength (RSS). For location estimation, angulation, lateration, and fingerprinting are the main techniques. AOA approaches require costly special antenna arrays and complex calculations while RSS and fingerprinting are too prone to changes in real-time. Therefore neither method is reliable nor highly accurate for indoor drone deployments. To avoid these pitfalls, RAIL uses trilateration and the TOA of the received ultrasound signals for localization. However, due to multi-path fading, accurately measuring TOA is challenging because of the copied version of the original signal. Thus, we first need to address multi-path fading before calculating the TOA of the received signal. To achieve that, RAIL uses a novel hybrid FH-CDMA communication scheme for its signal transmission.

\subsection{Implementation Challenges}
In terms of placement of the transmitter(s) and receiver(s) for a ranging-based localization, there are two general scenarios: either have the receiver(s) on-board the drone and keep the transmitter(s) in the room or the reverse. The localization calculation task takes place in the receiver side of the system. Therefore, if the receivers are not on-board the drone, another communication link for sending the final location estimation to the drone is required. This additional communication link can add latency and incur errors degrading the accuracy of the overall approach. Thus, we decided to place the receivers on the drone while keeping the transmitters in the room. Another observation is that having one transmitter in the room and multiple receivers on-board the drone raises several issues~\cite{Ultrasonic_Quadrotor_2019}. First, it adds extra weight to the drone and increases power consumption. Most importantly, due to the size of indoor drones, there is not enough space between the receivers, which induces additional errors.

To overcome these challenges, RAIL mounts only one receiver on-board the drone. We keep all the transmitters spatially distanced from each other in the room. However, this method raises yet a new challenge: the need for signal separation in the receiver. The receiver needs to be able to detect the TOA from individual transmitters. To rectify this, RAIL deploys a code division technique. It assigns a code to the transmitted signals of each of the ultrasound transmitters in the room. This means that the transmitted signals at each transmitter are encoded using a code that is orthogonal to all other transmitters' codes. Having four transmitters, RAIL generates a different orthogonal code for each transmitter using a Walsh-Hadamard matrix of size four. Data bits of each transmitter would be multiplied with one of the rows of this matrix. At the receiver side, received signals will be multiplied with all the four codes, and signals from each transmitter get detected.

\subsection{Hybrid FH-CDMA}
RAIL deploys a hybrid FH-CDMA technique for multi-path-robust ranging. To the best of our knowledge, this is the first time FH-CDMA has been used for localization. We show that FH-CDMA is the most desirable communication encoding to address both multi-path and signal separation problems. The hybrid FH-CDMA is a communication encoding that combines two well-known techniques, the Frequency Hopping (FH) and the Code Division Multiple Access (CDMA). RAIL uses this method to rectify the challenge of signal separation in the receiver with the multiple access capability and, at the same time, brings robustness against noise and the indoor multi-path fading using the frequency hopping technology.

In our system, we just need to guarantee that frequency hops occur fast enough that the transmission frequency has already been changed before the appearance of the first multi-path reflection at the receiver. If this holds, our receiver hops to another carrier frequency before the multi-path reflections can introduce errors. We select the hopping rate to be equal to the symbol rate which, according to the room channel characteristic, is fast enough to avoid multi-path and guarantee that our system is robust against multi-path fading. We show this assertion more clearly in the steps below.

First, we assign the same code to all the data bits coming from a specific ultrasound transmitter beacon in the room. Therefore, the symbols are no longer just a bit; they are coded bits that include four bits. Thus, data-symbols from different beacons are spread with their assigned code and generated coded symbols. Then, since the hop rate equals the symbol rate, each of those coded symbols is transmitted in different FH-channels. In our design, the coded symbols enable signal separation of different transmitters in the receiver side and the different frequency channels are used to manage the room's multi-path fading effect.

In our scheme, the transmitting signal of the $i$-th transmitter is modulated using Binary Phase Shift Keying (BPSK) modulation and then encoded with its dedicated code. We claim that since throughput is not essential in our scheme and precise detection matters the most, BPSK is the best option with its robust performance against noise and low error rate. Similar to \cite{Robust_Broadband}, the coded symbols are spread using a sinusoidal signal with a variable frequency depending on the pseudo-random code which is known both in the transmitter and receiver side:
\begin{eqnarray}\label{Tx_FHSSS}
s^{(i)}(t) = d^{(i)}\cdot c^{(i)} \cdot pT_B(t)\cdot \sin(2\pi f_mt+\phi),
\end{eqnarray}
where $T_B$ is the data symbol duration, $d^{(i)}\cdot c^{(i)}$ is the transmitted symbol of the $i$-th ultrasonic transmitter in the room where $d^{(i)}$ is the data bit and $c^{(i)}$ is the dedicated code to that transmitter, the rectangular pulse $pT_B$ is equal to $1$ for $0 \leq t < T_B$ and zero otherwise, and $f_m$ is the set of frequencies over which the signal hops. Then the received signal is in the form of:
\begin{eqnarray}\label{Rx_FHSS}
& r = \sum_{i=1}^{4} s^{(i)}(t-\tau_i) + \mathcal{M} + \mathcal{N},\nonumber
\end{eqnarray}
where $\tau_i$ is the propagation delay from the $i$-th transmitter to the receiver on-board drone that we are using for calculating the distance, $\mathcal{N}$ is the overall Gaussian noise, and $\mathcal{M}$ is the summation of all the multi-path fading effects:
\begin{eqnarray}\label{Multi-path}
\mathcal{M} =\sum_{i=1}^{4} \sum_{j=1}^{N} \alpha_{ij}\cdot s^{(i)}(t-\tau_{ij}),
\end{eqnarray}
where $\alpha_{ij}$ is the attenuation of path $j$ for the $i$-th transmitter and $\tau_{ij}$ is the time delay of the path $j$ for the $i$-th transmitter. We can defeat the multi-path fading effects, as long as we guarantee that hopping speed is faster than the time delay of each path ($\tau_j$). This will prevent interference from any of the reflected signals, because the receiver will change frequency by the time any multi-path signals can cause interference with the original signal. Having ensured that multi-path effects are eliminated using different FH-channels, the received signal would be only the delayed time of the transmitted signal plus noise:
\begin{eqnarray}
	r  = \sum_{i=1}^{4} s^{(i)}(t-\tau_i) + \mathcal{N}.
\end{eqnarray}
By multiplying the received signal in each code related to each transmitter, the received signal from the $i$-th transmitter in the receiver would be in the form of:
\begin{eqnarray}\label{Rx_FHSS_without_multipath}
r^{(i)}  = d^{(i)}\cdot pT_B(t-\tau)\cdot \sin(2\pi f_m(t-\tau)+\phi) + \mathcal{N}.
\end{eqnarray}
Therefore, by implementing a cross-correlation between the received signal and the known transmitted signal (the one without the time delay) and detecting the sample at which the peak occurs, the distance is calculated as the following:
\begin{eqnarray}\label{Distance_Calculation_2}
d = \frac{n_{samples}}{f_s}\cdot c_{sound},
\end{eqnarray}
where $n_{samples}$ is the sample number of the maximum peak, $f_s$ is the sampling frequency, and $c_{sound}$ is the speed of sound. 

\subsection{Three-dimensional Localization}
Having successfully measured the distance between an ultrasonic transmitter and the receiver, the next step is the three-dimensional localization of the receiver. In three-dimensional localization, to identify the location of a target object, we need to measure the distance between the target object and at least four distinct sources. Let's denote the distance between the receiver and the $i$-th transmitter as $d_i$. The position of the receiver is defined as $[x \ y \ z]^T$ (which is the position of the drone). Similarly, the position of the $i$-th transmitter is represented by $[x_i \ y_i \ z_i]^T$. Using trilateration rules we get: $(x_i-x)^2+(y_i-y)^2+(z_i-z)^2 = d_i^2$ for $i=1 \cdots n$.
We can then simplify these quadratic equations and write them down in the form of $\textbf{A}\textbf{x} = \textbf{b}$ where $\textbf{A}$ and $\textbf{b}$ are equal to:

\begin{small}
	\begin{eqnarray}
\textbf{A} & = & \begin{bmatrix}
2(x_n-x_1) & 2(y_n-y_1) & 2(z_n-z_1) \\
2(x_n-x_2) & 2(y_n-y_2) & 2(z_n-z_2)\\
\vdots     & \vdots     & \vdots\\
2(x_n-x_{n-1}) & 2(y_n-y_{n-1}) & 2(z_n-z_{n-1})\\
\end{bmatrix},\nonumber \\
\textbf{b} & = & \begin{bmatrix}
d_1^2 - d_n^2 - x_1^2 -y_1^2 -z_1^2 + x_n^2 + y_n^2 + z_n^2 \\
d_2^2 - d_n^2 - x_2^2 -y_2^2 -z_2^2 + x_n^2 + y_n^2 + z_2^2 \\
\vdots \\
d_{n-1}^2 - d_n^2 - x_{n-1}^2 -y_{n-1}^2 -z_{n-1}^2 + x_n^2 + y_n^2 + z_n^2
\end{bmatrix}.\nonumber
\end{eqnarray}
\end{small}
The vector $\textbf{x} = [x \ y \ z]^T$ which includes the coordinate of the target drone would be: $\textbf{x} = (\textbf{A}^T\textbf{A})^{-1}\textbf{A}^T\textbf{b}$. 

\section{Simulation Analysis} \label{Simulation}
This section evaluates the performance of the proposed FH-CDMA localization and presents a benchmark for our experimental tests. First, the simulation setup is described in \ref{Simulation_Setup}, and then the results of our simulation are depicted in \ref{Simulation_Results}.

\subsection{Simulation Setup} \label{Simulation_Setup}
The performance of the localization scheme proposed in section \ref{Robust FH-CDMA Localization} is assessed by simulation in MATLAB. Similar to \cite{ROLATIN}, we locate the transmitters at the $(x,y,z)$ coordinates equal to $(2.5,0,1.5)$, $(5,2.5,2.5)$, $(2.5,5,2)$, and $(0,5,3)$ where all the numbers are in the meter unit. To better observe how the simulation is conducted, we divide it into three sub-systems, as shown below.

The transmitter sub-system, which is the ultrasonic transmitters at the known positions in the room, generates the desired FH-CDMA signals. We used signals in the frequency range of $20$~KHz to $50$~KHz because of two reasons. First, to avoid inciting excessive audible noise or facing interference from human-generated voice, we pick frequencies over $20$~KHz to prevent overlapping with the audible frequency range. On the other hand, according to the Nyquist theorem, the sampling rate needs to be at least twice the maximum frequency to avoid aliasing; hence, if the system works in the frequency range of $20$~KHz to $50$~KHz, then the sampling frequency needs to be at least $100$~KHz. To avoid the cost of processing and equipment, dealing with high frequency is not suitable; therefore, we do not transmit above $50$~KHz, which means that the sampling rate could be $100$~KHz or more. 
To generate the FH-CDMA waveform successfully, we use $6$ different frequency hops with $5$~KHz bandwidth dedicated to each hop. Also, it assigns a code to each transmitter, so every data bit of each transmitter would first be multiplied with the code and then transmitted via one of the six frequency hops centered at frequencies $22.5$~KHz, $27.5$~KHz, $32.5$~KHz, $37.5$~KHz, $42.5$~KHz, and $47.5$~KHz. The codes are orthogonal to each other and made by a Walsh-Hadamard matrix of size $4$. At each hop, one data symbol which already been multiplied by its code would be transmitted, so the hop rate is equal to the data bit rate (actual data bit rate before multiplying by the code) which is fast enough to mitigate the multi-path fading effect of the indoor environment. Although a sampling rate of $100$~KHz would be enough for our simulation, we picked sampling frequency ($f_s$) equal to $340$~KHz to ensure it would be large enough to avoid aliasing, and it also helps to simplify some of our calculations. Since the throughput is not essential in our case, we use BPSK modulation, which does not have a high transmission rate, but is highly robust against noise.

The channel sub-system is used to add white Gaussian noise (AWGN) and simulate the multi-path fading of the indoor environment. The drone’s movement is assumed to be restricted to a rectangular room with dimensions of $5$~m~$\times$~$5$~m~$\times$~$4$~m with all the regular office environment considerations. We use a Rayleigh channel with several paths for simulating the multi-path fading effect due to the reflection of the signal from walls, floor, and ceiling. The Rayleigh channel parameters that we set for our simulations are the sample rate, maximum Doppler-shift, number of different paths, delay of each path, and average path gain. We put all these parameters concerning a typical indoor room environment using image theory.     

Finally, the receiver sub-system, which is the ultrasound receiver on-board the drone, separates the signals from different transmitters by multiplying them into the transmitters' codes and demodulating the frequency hopped signals. Then, it cross-correlates the received signal with the original version and finds the bit which makes the peak in the  cross-correlation, and using that, it estimates the distance from each transmitter to the drone using: 
$d = n_{samples}\times c_{sound}/f_s$, where $n_{samples}$ is the sample number that the maximum cross-correlation occurs and $f_s$ is the sampling frequency.

\subsection{Simulation Results} \label{Simulation_Results}
We assess the performance of the FH-CDMA localization by calculating the error between the actual position of the drone and our estimated position. A Monte Carlo method with an adequately large number of iterations is used for each of the simulations.

\begin{figure}
	\centering
	\includegraphics[width=\linewidth]{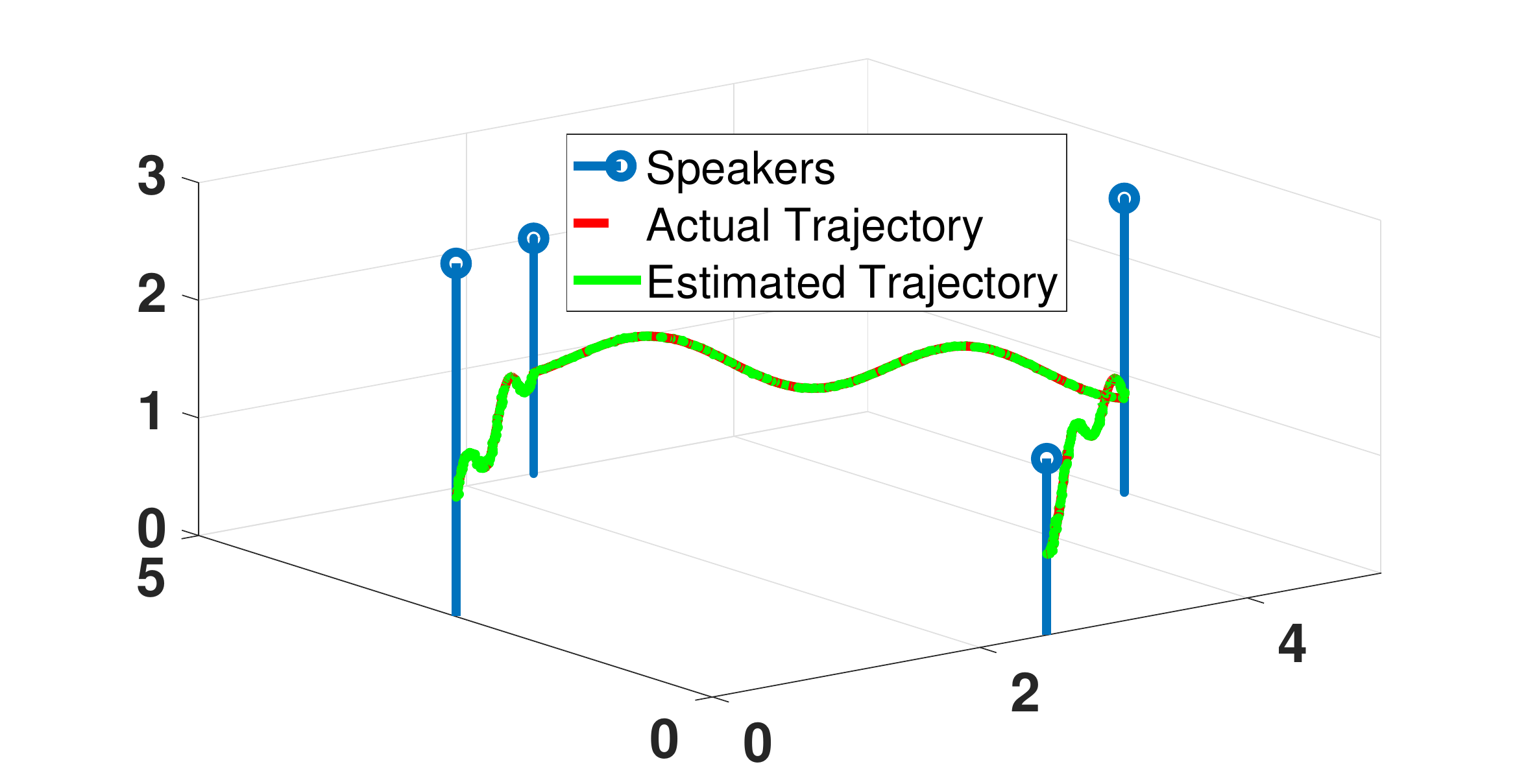} 
	\caption{Representation of the ultrasound speaker transmitter placement in the room and a comparison between the estimated trajectory of a drone and its actual trajectory.}
	\label{trajectory}
\end{figure}

In Fig.~\ref{trajectory}, a drone's actual trajectory, as well as the estimated trajectory using just FH-CDMA localization, is shown. The locations of the ultrasound speaker transmitters are also indicated in this figure. The actual and estimated trajectories seem to overlap perfectly because the localization estimation error is significantly small relative to the room's dimensions.

\begin{figure}
	\centering
	\includegraphics[width=\linewidth]{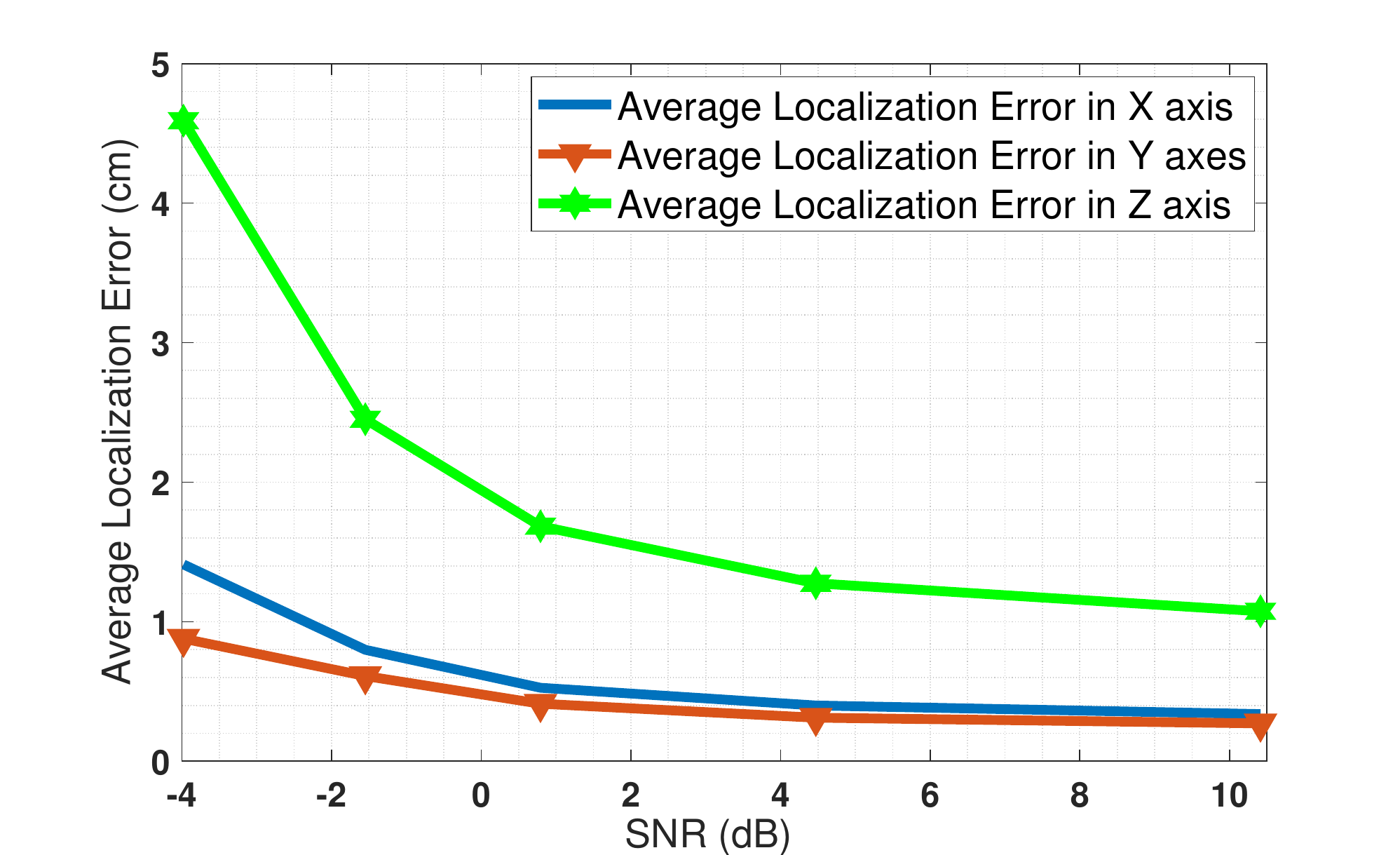} 
	\caption{Average Localization error vs. SNR (dB).}
	\label{SNR}
	\vspace{-0.15in}
\end{figure}

Fig.~\ref{SNR} shows the relationship between FH-CDMA localization performance and the signal-to-noise ratio (SNR) of the signal received by the ultrasound receiver. The localization error is inversely proportional to the SNR value of the signal, as expected. In the figure, note that the $Z$-axis localization error is much greater than that of the $X$ or $Y$ axis at any given SNR.

By conducting more simulations with different drone trajectories, we observed that the error of $Z$-axis localization is always drastically more than the $X-Y$ plane localization error. This is despite the fact that all the X, Y, and Z axes should have similar errors because they face a similar channel.

\section{Experimental Results and Evaluation} \label{Experimental Tests and Results}

\subsection{Additional Sensor for Height Estimation}
To improve the height estimation, RAIL deploys an additional ultrasonic transceiver mounted on-board the drone to estimate the height continuously. Then, using a filter, it incorporates this measurement with the $Z$-axis estimation that is already available from the previous step. This significantly improves the $Z$-axis estimation accuracy.

The ultrasonic transceiver is located on-board the drone facing upwards and finds the distance between the drone and the ceiling by calculating the time of flight of the ultrasonic signal transmitted from the sensor, after it is reflected from the ceiling. Then, simply by subtracting this result from the room's height, the drone's height at each moment is calculated. The channel between the drone and the ceiling is usually more reliable than the one between the drone and the floor because usually, there are no objects between the drone and the ceiling that induce errors. Following shows the height estimation using this extra ultrasonic transceiver:
\begin{eqnarray}
	d = \ c_{sound}\cdot t/2 \ ; \ h_{drone} = \ H - d, \nonumber
\end{eqnarray}
where $d$ is the distance between the drone and the ceiling, $t$ is the total time that takes the signal to travel from ultrasonic transceiver on-board the drone and hit the ceiling, reflecting, and is received in the ultrasonic transceiver on-board the drone, $H$ is the room height, and $h_{drone}$ is the estimation for drone's height.

\subsection{Experimental Setup}
\begin{figure}
	\centering
	\includegraphics[height=1.1in,width=2.25in,trim={3cm 21cm 19cm 1cm},clip]{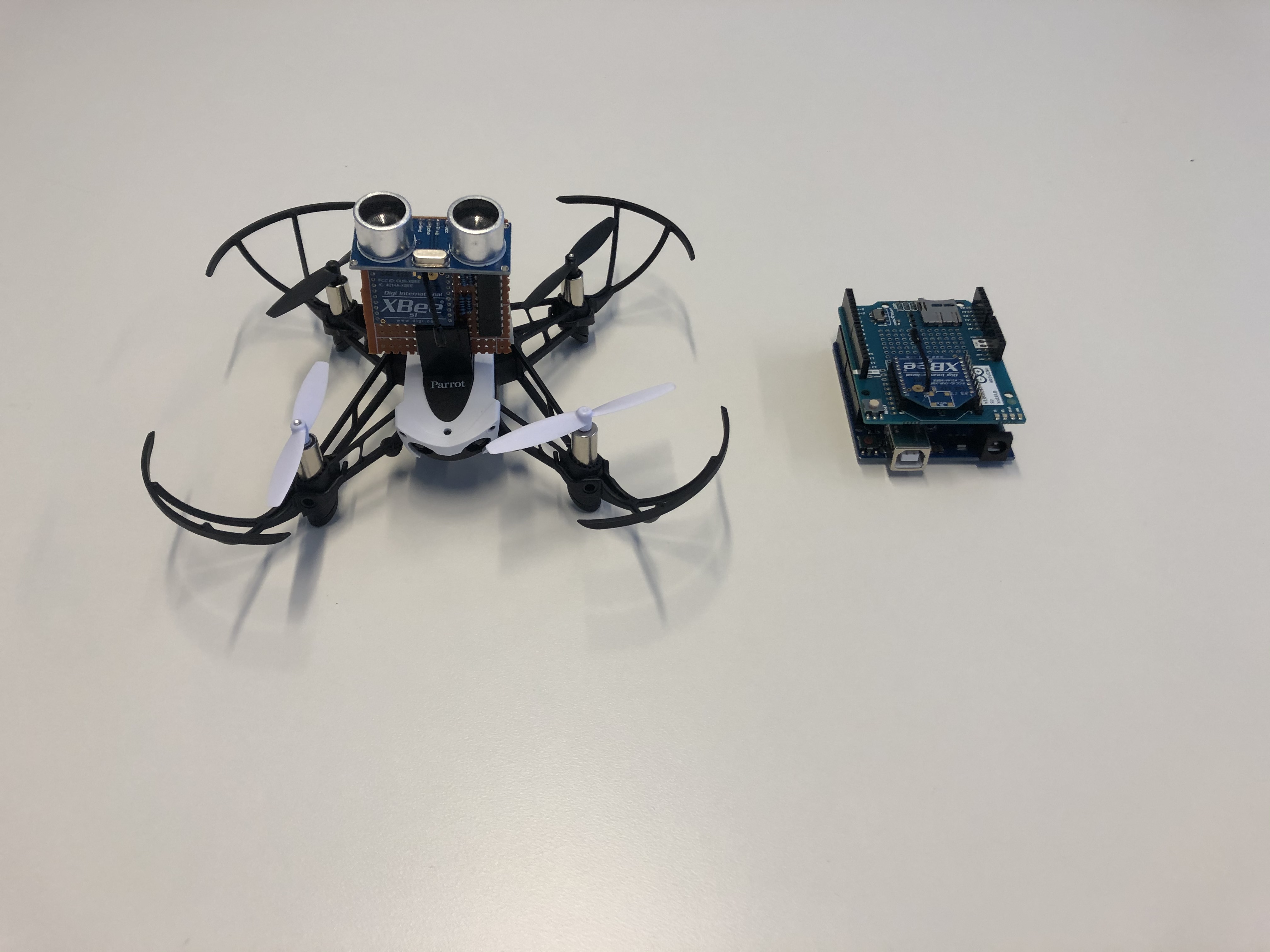} 
	\caption{Parrot Mambo Drone equipped with the ultrasound transceiver system at the left and the receiver side on the right part.}
	\label{testbed}
\end{figure}

We conducted different experimental tests based on and coupled with our MATLAB simulations. As depicted in Fig.~\ref{testbed}, the experimental test-setup consists of two stations: first, the drone and the system on-board it, and the second one is the ground control station which helps to input the transmitted data into the MATLAB program running on a Dell XPS $15$ laptop. The drone is being used for the experiment is a Parrot Mambo Drone. It is a cheap, off-the-shelf, and ultralight drone suitable for indoor experiments, and also it has the capability of carrying some light loads. The designed system mounted on-board the drone consists of an Arduino Uno micro-controller connected to an $HC$-$SR04$ sensor for ultrasonic distance measurement purposes and a XBee S1 module for wireless communication with the ground controller. In the ground control unit, another Arduino Uno micro-controller connected to a XBee S1 receives the data and transfers it into the MATLAB program running on the laptop. All the experiments are conducted in a hallway inside the building with dimensions $5$~m$~\times$~$5$~m~$\times$~$4$~m.

\subsection{Overall Combined Results}
\begin{figure}
	\centering
	\begin{subfigure}{.42\textwidth}
		\centering
		\includegraphics[width=\textwidth]{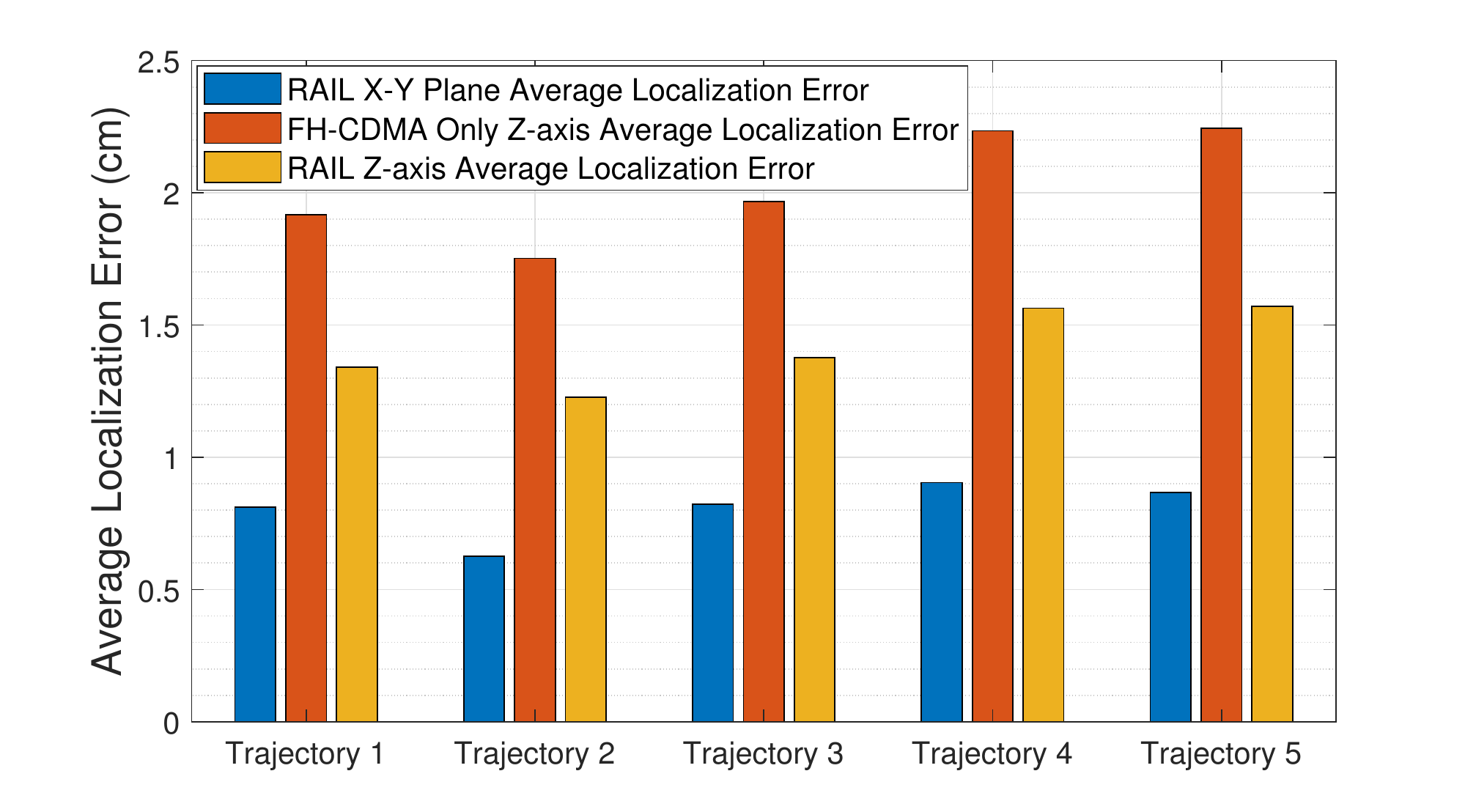}
		\caption{}
		\label{Experiment_XYvsZ}
	\end{subfigure}
	\begin{subfigure}{.42\textwidth}
		\centering
		\includegraphics[width=\textwidth]{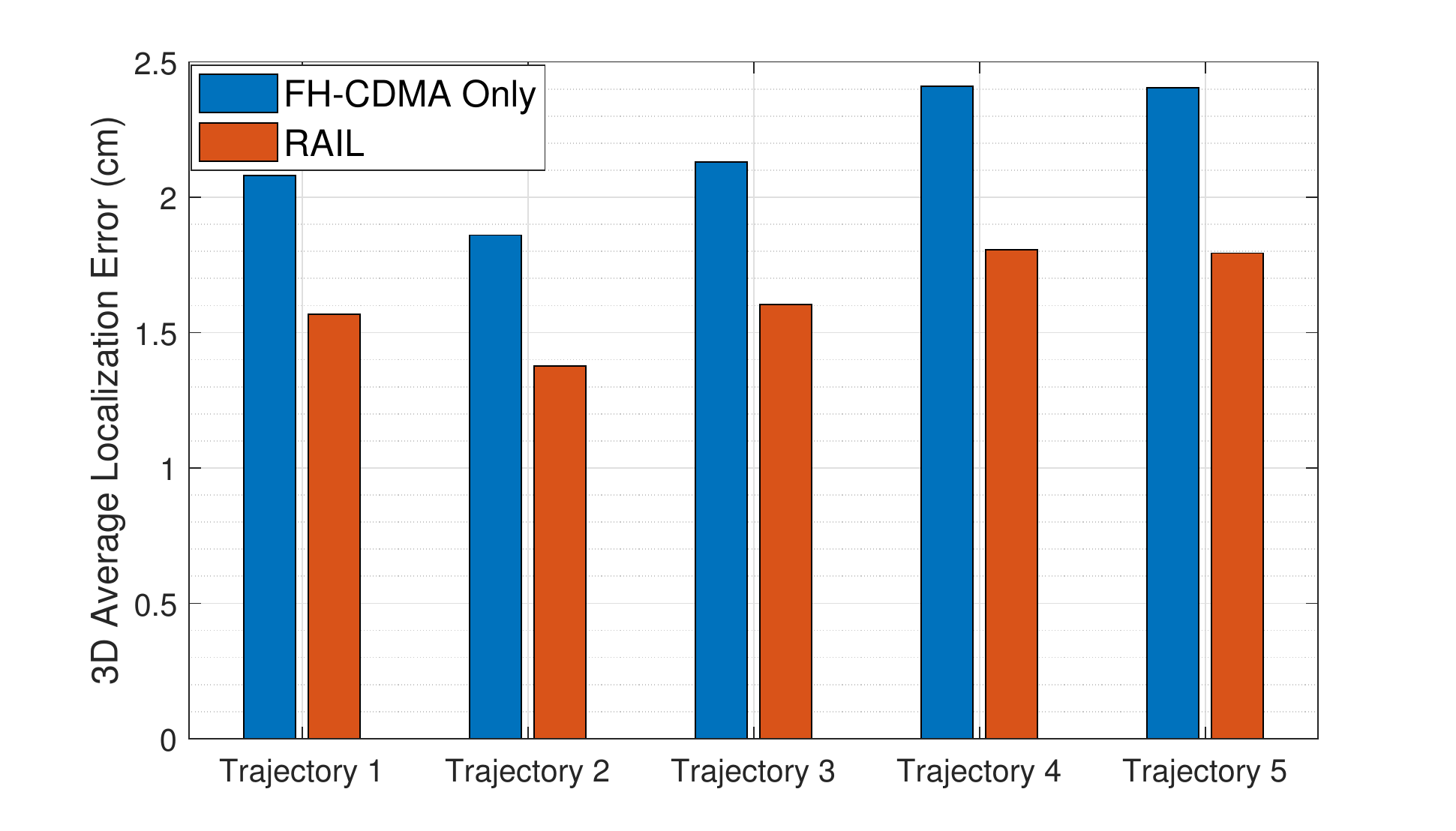}
		\caption{}
		\label{Experiment_Overall}
	\end{subfigure}
	\caption{(a) Evaluating the performance of RAIL: Comparison between the $X-Y$ plane average estimation error and the $Z$-axis. (b) Assessing the performance of RAIL: overall three-dimensional localization accuracy. In both figures, we plot the comparison for just $5$ random trajectories out of many more to avoid having cluttered figures.}
	\label{}
\end{figure}

In Fig.~\ref{Experiment_XYvsZ}, the localization error of the $Z$-axis with the $X-Y$ plane is compared. This figure justifies the necessity of having the auxiliary sensor for height estimation. Furthermore, this figure shows how the last step of RAIL further improves the $Z$-axis estimation by constantly transferring the measured data from the ultrasound sensor on-board the drone ($HC$-$SR04$) to the receiver module connected to the Dell XPS $15$ laptop. Therefore, RAIL successfully improves the $Z$-axis estimation.

In Fig.~\ref{Experiment_Overall}, the performance of RAIL with that of the benchmark scheme (which relies only on FH-CDMA distance estimation to localize a target drone) in terms of the overall three-dimensional localization error is compared. The average value of three-dimensional localization error for RAIL is $1.5$~cm. As is seen in the figure, the benchmark scheme’s localization error is almost twice that of RAIL. This is because the benchmark scheme merely focuses on mitigating ranging-based error by deploying the FH-CDMA communication scheme for localization. Other drone localization schemes proposed in the literature do the same and try to improve the localization accuracy by offering their technique to mitigate the ranging-based error. However, RAIL proposes a scheme that deals with both ranging-based errors and fixes the $Z-$axis estimation error induced by relative geometry between the transmitters and the receiver and further improves the accuracy.

To report an overall average localization error for RAIL in any possible scenario and compare it with the state-of-the-art, we conducted ample simulation and experimental tests with different trajectories with random paths and in various environments. Based on our evaluations and the overall report of the other work in the literature~\cite{ROLATIN, Robust_Broadband, Ultrasonic_Quadrotor_2019, Spread_Spectrum_and_MEMS, Spread_Spectrum_Ultrasound_and_Time-of-Flight_Cameras}, RAIL achieves significant improvement in comparison with the previous drone localization schemes. For instance, in \cite{ROLATIN}, their approach incurs a high $Z$-axis estimation error, and they did not propose any solution to fix it. Moreover, their scheme requires an additional communication link which may induce more latency and error. We rectify this issue and eliminate the extra link by changing the setup and leveraging the CDMA technique to provide signal separation at the receiver. On the other hand, \cite{Spread_Spectrum_and_MEMS} has a localization error of at least $2$~cm only for two dimensions without even addressing three-dimensional localization. In \cite{Ultrasonic_Quadrotor_2019}, the proposed scheme has an average error of $5.2$~cm for three-dimensional localization for drones which is more than three times what RAIL achieves.

\section{Conclusions} \label{Conclusion}
In this paper, we designed and evaluated RAIL, a novel three-dimensional localization approach for drones in GPS-denied environments. To our knowledge, RAIL is the first scheme to leverage the hybrid \textit{FH-CDMA} for multi-path resilient ranging and provide high-accuracy three-dimensional localization, which is necessary for successful autonomous drone navigation. We evaluated RAIL over an ample of different random trajectories. Our simulation and experimental results demonstrate that RAIL's localization error is $1.5$ centimeters on average in three-dimensional space, making RAIL an excellent alternative to GPS for any indoor drone deployment case.

\footnotesize
\bibliographystyle{IEEEtran}
\bibliography{ref_RAIL_arXiv_v0}

\end{document}